# "A new metric of consensus for Likert scales"

Claveria O

## Abstract

In this study we present a metric of consensus for Likert-type scales. The measure gives the level of agreement as the percentage of consensus among respondents. The proposed framework allows to design a positional indicator that gives the degree of agreement for each item and for any given number of reply options. In order to assess the performance of the proposed metric of consensus, in an iterated one-period ahead forecasting experiment we test whether the inclusion of the degree of agreement in consumers' expectations regarding the evolution of unemployment improves out-of-sample forecast accuracy in eight European countries. We find evidence that the degree of agreement among consumers contains useful information to predict unemployment rates in most countries. The obtained results show the usefulness of consensus-based metrics to track the evolution of economic variables.

Oscar Claveria AQR-IREA, University of Barcelona (UB). Tel.: +34-934021825; Fax.: +34-934021821. Department of Econometrics, Statistics and Applied Economics, University of Barcelona, Diagonal 690, 08034 Barcelona, Spain. E-mail address: oclaveria@ub.edu

**Acknowledgements**

This research was supported by by the projects ECO2016-75805-R from the Spanish Ministry of Economy and Competitiveness.

## 1. Introduction

The measurement of consensus is a key component of data analysis and the study of human behaviour (von der Gracht, 2012; Wirth, 1948). Likert scales were developed by Likert (1932) to measure people's attitudes. It is the most common approach to scaling responses in survey research. Likert scales result when survey participants are asked to rank their agreement with a set of items on a scale that has a limited number of possible responses that are presented in a sequence. The number of responses is usually five, which can take the following form: "strongly agree", "agree", "undecided", "disagree", "strongly disagree". The number of responses may vary. See Cox (1980) for a discussion on the optimal number of response alternatives.

Likert scales are very useful in opinion surveys. Economic tendency surveys, which are conducted among businesses and consumers, use Likert-type questionnaires with three and five reply options respectively. These surveys are the main source for eliciting the expectations of economic agents. As a result, survey-based expectations are widely used for the construction of economic indicators (Claveria et al., 2017), as explanatory variables in quantitative forecasting models (Kaufmann and Scheufele, 2017) and also to test economic hypothesis (Binder, 2015).

In this study we present a measure of consensus among respondents in Likert-type questionnaires. The proposed metric is derived by means of a conceptual framework that conveys a geometric interpretation. The consensus measure has a self-explanatory interpretation, as it provides the percentage of agreement among respondents. In order to assess the performance of the proposed statistic, we use it to compute the level of consensus between consumers regarding their expectations about unemployment in eight European countries. We then scale the metric and evaluate whether it helps to improve the accuracy of unemployment rate forecasts.

With this objective, we design an iterated one-period ahead forecasting experiment in which we generate out-of-sample predictions of the unemployment rates using an autoregressive model as a benchmark. We then replicate the experiment including a consensus-based indicator as a predictor, and test if the reduction in the mean percentage absolute forecast error is statistically significant.

## 2. Methodology

In this section we present a methodology to compute a metric of consensus among survey respondents of Likert-type questionnaires. The approach is based on a geometric framework applied to proxy economic uncertainty (Claveria et al., 2018) and to determine the likelihood of disagreement among election outcomes (Saari, 2008).

Let us assume a Likert-type questionnaire with $N$ reply options, where $R_{i,t}$ denotes the aggregate percentage of responses in category $i$ at time $t$, where $i = 1, \ldots, N$. As the sum of the reply options adds to 100, a natural representation of the vector $X_t$ containing all the information from the surveyed units at a given time $t$ is as a point on a simplex (Coxeter, 1969).

The interior of this simplex encompasses all possible combinations of reply options, which correspond to the barycentric coordinates of each point in time. Each of the $N$ vertexes corresponds to a point of maximum consensus. We propose measuring the level of agreement as the ratio between the distance of the point to the barycentre and that form the barycentre to the nearest vertex. Hence, the measure of conensus at a given period $t$ can be formalised as:

$$C_t = \frac{\sqrt{\sum_{i=1}^{N}\left(R_{i,t} - \frac{100}{N}\right)^2}}{\sqrt{(N-1)/N}} \qquad (1)$$

This metric reaches the maximum of one hundred percent when a response category draws all the responses, and the minimum value of zero when the answers are equidistributed among the $N$ response categories. This measure incorporates the share of neutral responses and allows to capture the trajectories of the different states.

## 3. Empirical results

The empirical analysis is done for consumer survey data regarding unemployment expectations. We use monthly data from the joint harmonised EU consumer survey conducted by the European Commission (https://ec.europa.eu/info/business-economy-euro/indicators-statistics/economic-databases/business-and-consumer-surveys_en). The sample period goes from January 2007 to December 2017.

Consumers are asked whether they expect a certain variable to "sharply increase", "slightly increase", "remain constant", "slightly decrease" or "sharply decrease". Thus, we respectively denote the aggregated percentages of the individual replies in each category as $PP_t$, $P_t$, $E_t$, $M_t$ and $MM_t$. Consumers are also faced with questions with three reply options, in which they are asked whether they expect a variable to "increase", "remain constant" or "decrease". In this case, the percentages of respondents are respectively noted as $P_t$, $E_t$, and $M_t$.

In Fig. 1 we represent the resulting simplexes that explicitly incorporates all the components for both three and five reply options respectively.

**Fig. 1.** Simplex – Equilateral triangle vs. Regular pentagon

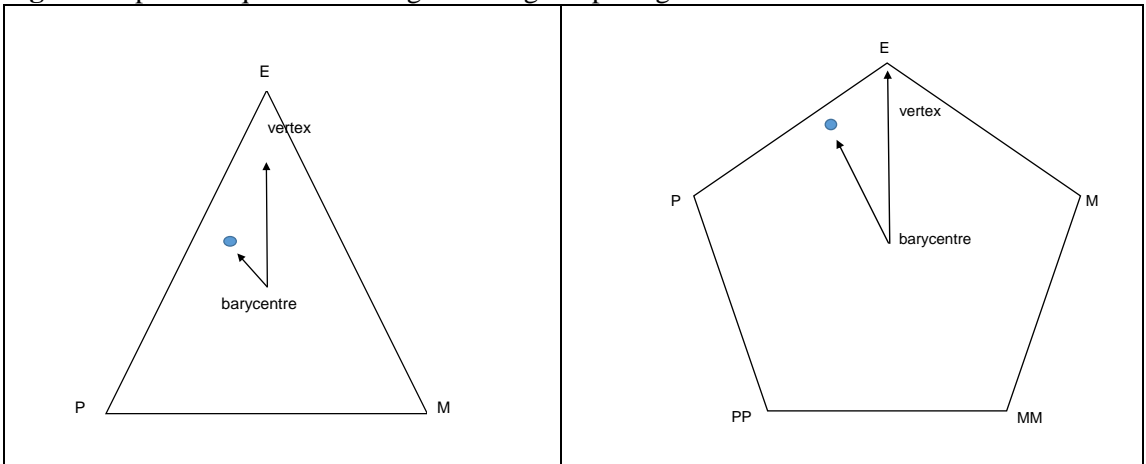

1. Note: The equilateral triangle corresponds to the three-reply option, where are E denotes the % of "remains constant" replies, P the % of "increase", and M the % of "decrease". The regular pentagon corresponds to the five-reply question, where E denotes the % of "remains constant" replies, P the % of "slight increase", PP the % of "sharp increase", MM the % of "sharp fall", and M the % of "slight fall". The grey point in the simplex corresponds to a unique convex combination of all reply options for a given period in time.

In this study we compare the performance of the proposed metric for both scenarios: the consensus for three reply options ($C3_t$) and for five ($C5_t$). As unemployment expectations are elicited via five reply options, in order to compute $C3_t$ we opt for grouping all positive responses in $P_t$, all negative ones in $M_t$, and incorporating the "do not know" share in $E_t$.

In Fig. 2 we graph the evolution of both of both consensus measures using as a backdrop the distribution of survey responses grouped in three categories during the sample period.

**Fig. 2.** Evolution of the distribution of survey responses by category and the consensus metrics

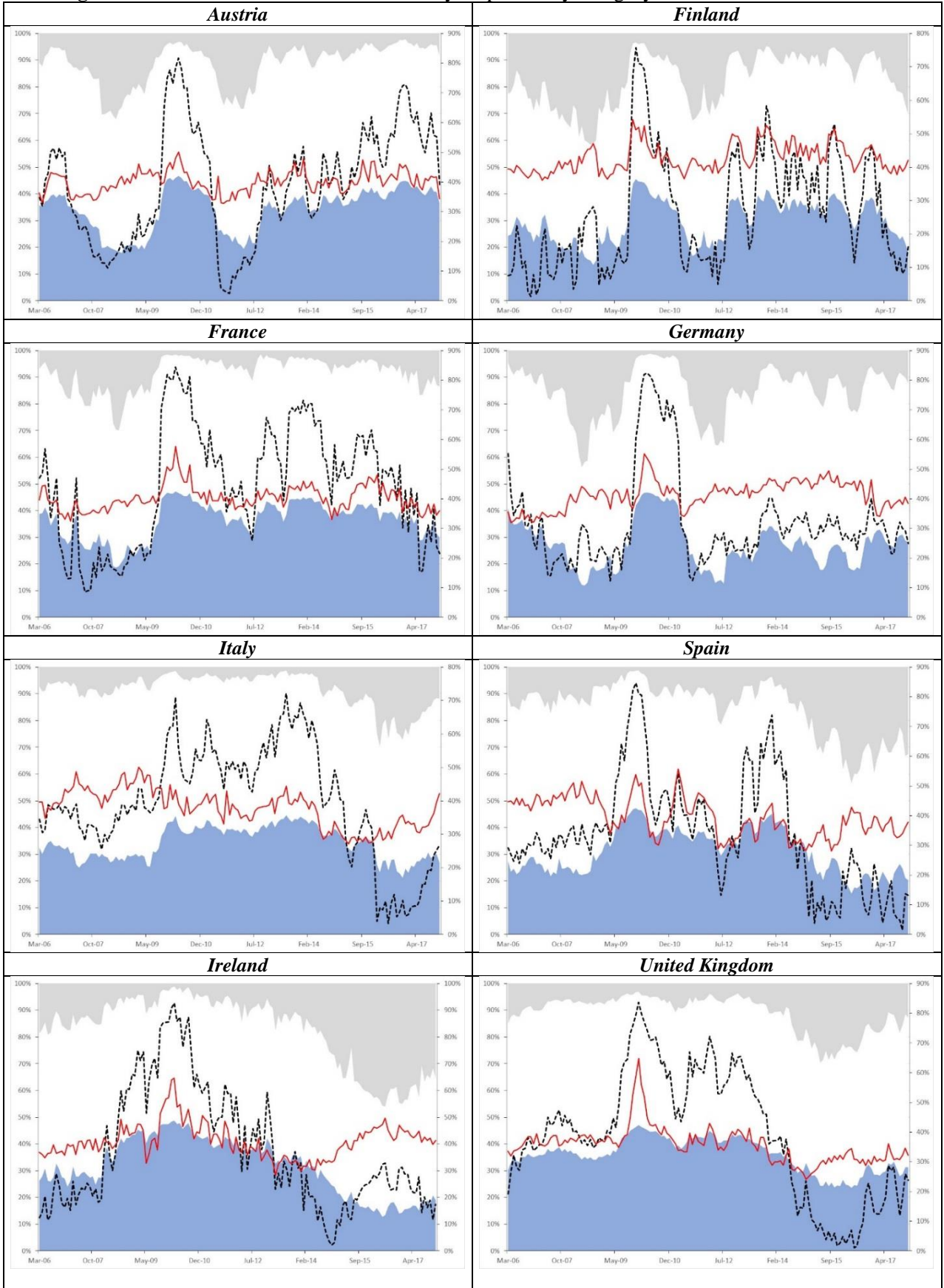

2. Note: The blue area represents the evolution of the percentage of "fall" responses (sharply and slightly) regarding the level of unemployment over the next 12 months, while the light grey area represents the % of "increase" responses (sharply and slightly) and the white area the % of "remain the same (no-change)" responses. The black dashed line represents the evolution of the of the consensus measure for three categories, while the red line the consensus for five categories.

We can see that both measures co-evolve during the sample period, being the five-response consensus metric ($C5_t$) the one that shows less dispersion. This notion is further confirmed in Table 1, where we present the results of the simulated sampling distributions of the simplexes defined in Fig. 1 by generating a uniform set of 10,000 points.

**Table 1.** Summary statistics of simulated distribution of consensus metrics

|  | Mean | Std. Dev. | Min. | Max. | Range | IQR |
|---|---|---|---|---|---|---|
| $C3_t$ | 33.797 | 17.797 | 0.162 | 99.337 | 99.175 | 25.619 |
| $C5_t$ | 35.672 | 13.377 | 1.581 | 98.755 | 97.174 | 17.477 |

Note: The range is obtained as the difference between the maximum and the minimum values of the distribution. The IQR refers to the interquartile range, which is obtained as the difference between upper and lower quartiles, Q3−Q1.

To test if the proposed measure of consensus helps to improve the accuracy of unemployment forecasts, we generate two proxies of the unemployment rate by smoothing both metrics with a simple moving average and scaling them by means of a rolling regression. The resulting consensus-based proxies are respectively denoted as $SC3_t$ and $SC5_t$. We use seasonally adjusted unemployment rates provided by Eurostat (http://ec.europa.eu/eurostat/web/lfs/data/database). Then, in order to evaluate their forecasting performance, we introduce them as explanatory variables in autoregressive (AR) models. This model is usually referred to as ARX or dynamic regression model. Regarding the number of lags that should be included for each period in every country, we choose between models with a minimum of 1 lag and a maximum of 12 lags, selecting the model with the lowest Akaike's information criterion (Akaike, 1974).

We design an iterated one-period ahead forecasting experiment in order to assess the forecast accuracy of the out-of-sample predictions. First, we use the last twelve periods to compute the mean absolute percentage forecast error (MAPFE), which is a scale-independent measure. The fact that we are dealing with positive data and comparing countries with different unemployment rates, makes the MAPE particularly suitable in this case (Hyndman and Koehler, 2006). Second, we run the Diebold-Mariano (DM) test of forecast accuracy (Diebold and Mariano, 1995) to evaluate whether the inclusion of the consensus-based proxies significantly lower forecast errors.

In Table 2 we present the results of the out-of-sample forecasting evaluation. We find that the consensus-based proxies significantly improve the AR-generated forecasts in Finland, France, Italy, Spain and Ireland. These results are in line with those obtained by Claveria (2018) and allow us to conclude that in most cases, the information coming from the degree of consensus significantly helps to improve forecast accuracy of unemployment rate predictions.

**Table 2.** Out-of-sample forecast accuracy – MAPFE values (2017.01-2017.12)

|                | AR    | ARX with $SC3_t$ | ARX with $SC5_t$ |
|----------------|-------|------------------|------------------|
| Austria        | 0.659 | 0.716            | 0.633            |
|                |       | (0.251)          | (0.599)          |
| Finland        | 0.635 | 0.452            | 0.457            |
|                |       | (3.524*)         | (2.641*)         |
| France         | 1.273 | 1.089            | 1.171            |
|                |       | (5.158*)         | (0.803)          |
| Germany        | 0.734 | 0.672            | 0.748            |
|                |       | (0.465)          | (-0.591)         |
| Italy          | 1.102 | 0.989            | 0.949            |
|                |       | (12.904*)        | (2.451*)         |
| Spain          | 1.579 | 1.155            | 0.754            |
|                |       | (5.418*)         | (4.224*)         |
| Ireland        | 2.404 | 1.664            | 1.263            |
|                |       | (4.218*)         | (10.688*)        |
| United Kingdom | 0.614 | 0.597            | 0.638            |
|                |       | (0.586)          | (-0.337)         |

Note: MAPFE stands for the mean percentage absolute forecast error. Diebold-Mariano test statistic with NW estimator between brackets. Null hypothesis: the difference between the two competing series (AR vs. ARX) is non-significant. A negative sign of the statistic implies that the second model has bigger forecast errors. * Significant at the 5% level.

## 4. Conclusion

This paper presents a novel measure of consensus for Likert-type questionnaires. The proposed metric provides the percentage of agreement for each item independently of the number of reply options, and incorporating the information coming from the share of neutral responses.

With the aim of assessing the performance of the proposed metric of consensus, we have used it compute the level of consensus in consumers' unemployment expectations in eight European countries.

We have designed a two-step iterated forecasting experiment in which we first generate out-of-sample predictions of the unemployment rates using an autoregressive model as a benchmark, and then we replicate the experiment including a consensus-based proxy of unemployment as a predictor so as to test whether it significantly improves forecast accuracy.

We find that the degree of agreement improves forecast accuracy in all countries. The reduction in forecast errors is statistically significant in five of the countries included in the analysis (Finland, France, Ireland, Italy and Spain). On the one hand, this finding reveals that the degree of agreement in consumers' expectations contains useful information to predict unemployment rates. On the other hand, these results underline the importance of the measurement of consensus, and hint at the usefulness of consensus-based metrics to track the evolution of economic variables.

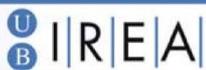

Institut de Recerca en Economia Aplicada Regional i Públic
*Research Institute of Applied Economics*

**WEBSITE:** www.ub-irea.com • **CONTACT:** irea@ub.edu

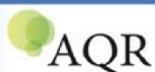

Grup de Recerca Anàlisi Quantitativa Regional
*Regional Quantitative Analysis Research Group*

**WEBSITE:** www.ub.edu/aqr/ • **CONTACT:** aqr@ub.edu

**Universitat de Barcelona**
Av. Diagonal, 690 • 08034 Barcelona